# High Isolation Improvement in a Compact UWB MIMO Antenna

Hossein Babashah[a], Hamid Reza Hassani[a], Sajad Mohammad-Ali-Nezhad[b*]

[a]Electrical and electronics Engineering Department, Shahed University , Tehran, Iran

[b]Electrical and electronics Engineering Department, University of Qom , Qom, Iran

[*]Corresponding author: Sajad Mohammad-Ali-Nezhad, Email: s.mohammadalinezhad@qom.ac.ir

A compact multiple-input-multiple-output (MIMO) antenna with very high isolation is proposed for ultrawideband (UWB) applications. The antenna with a compact size of $30.1 \times 20.5\ mm^2$ ($0.31\lambda_0 \times 0.21\lambda_0$) consists of two planar-monopole antenna elements. It is found that isolation of more than $25\ dB$ can be achieved between two parallel monopole antenna elements. For the low frequency isolation, an efficient technique of bending the feed-line and applying a new protruded ground is introduced. To increase isolation, a design based on suppressing surface wave, near-field, and far-field coupling is applied. The simulation and measurement results of the proposed antenna with good agreement are presented and show a bandwidth with $S_{11} \leq -10\ dB$, $S_{12} \leq -25\ dB$ ranged from 3.1 to 10.6 $GHz$ making the proposed antenna a good candidate for UWB MIMO systems.

## INTRODUCTION

MIMO technology has aroused interest because of its application in 4G, RFID, Digital Home and WLAN. Demand for high data rate and as a result huge bandwidth is increasing. In 2002 US-FCC approved unlicensed use of 3.1- 10.6 GHz frequency band at low energy level [1], therefore in order to improve capacity of the system, ultra wideband MIMO antenna has been developed for commercial systems . UWB MIMO antenna with high isolation has application in automotive communications and radar imaging systems [2,3].

When several antennas come out in close proximity, they suffer from severe mutual coupling, which results in lower antenna efficiency and loss of bandwidth and further degrade the performance of either diversity gain or spatial multiplexing schemes [3]. So the question then arises as to how to put together antenna elements with low coupling and occupying the least possible space. Because these two properties contradict each other, the problem is very challenging. The mutual coupling is also attributed to three phenomena such as near-field coupling, far-field coupling and surface wave coupling [4].

Many techniques and MIMO structures have been proposed for compact MIMO systems. In [5–7] the size of the reported antennas are large and may not be suitable for the present portable applications. In [3,6] and [8-10] the antenna elements are perpendicular to each other.

Certain techniques are also reported to improve isolation. Methods include using simple and fractal based DGS [11], EBG [12], soft surface structures [13], and meander line slots in between the antenna elements [14], etc.

Among the aforementioned designs none of them could achieve a very high isolation in such a small size at low frequency levels because in small size structures reducing mutual coupling at these frequencies due to long wavelength is very challenging.

In this paper, a new technique to improve isolation between two closely-spaced UWB MIMO antenna elements is introduced and an UWB MIMO antenna with a compact size of 30.1×20.5 $mm^2$ is presented. With the proposed technique, the three causes of mutual coupling are obliterated and a high isolation at a very small size is achieved. In the present work, the antenna elements are placed in parallel to each other in which case achieving a high isolation is much more difficult than being perpendicular due to polarization. A long strip is placed in between the two antenna elements to reduce mutual coupling and the ground of two elements are separated from each other to enhance isolation at low frequency levels. Also, through changing the feed structure and applying a new protruded ground along the feed line, which can be considered as part of radiator, the isolation is further improved. A prototype of the proposed antenna is fabricated, and measured results are compared with those of simulation. The proposed antenna shows several advantages such as small size, very high isolation while having parallel elements and good efficiency. The simulations are carried out using the commercially available software package ANSYS HFSS.

**Antenna Design**

The geometry of the proposed UWB MIMO antenna is shown in Figure 1. A planar rectangular shaped antenna is designed with size of $10 \times 8\ mm^2$. The antenna is printed on a Rogers substrate, RT/duroid 5880 with a relative permittivity 2.2, a loss tangent 0.0009 and a thickness of $0.787\ mm$. The antenna consists of two conventional symmetric planar-monopole antenna elements (PM1 and PM2). The two antenna elements are placed in parallel with each other and fed through a microstrip feed line. The two rectangular-shaped identical radiators are printed on the front side of the substrate and have dimension $L_r$ and $W_r$.

Printed UWB monopole antenna can provide wide lobe pattern, if such antenna in a MIMO system is designed to have a high gain and low mutual coupling then the system can act as a MIMO UWB diversity antenna [3]. To enhance isolation each of the radiators are fed by a $50\ Ohm$ bent microstrip line which changes the current distribution between antenna elements and suppresses near-field and far-field coupling. A common long strip is placed in the middle of the ground plane with dimension of $W_s \times L_s$ to reduce the mutual coupling by resonating. For better impedance matching at the higher frequencies, a small rectangular strip with size of $W_{s1} \times L_{s1}$ is printed on the upper edge of feed line. Two h-shaped symmetric slots are protruded in the ground to engineer field pattern causing mutual coupling. This reduction happens because the electromagnetic wave experience discontinuity along its way of propagation which also results in radiation [15]. Two symmetric short strips are also etched in the ground to improve impedance matching. To attain high isolation at the higher frequencies, the grounds are separated by protruding the common ground with dimensions of $L_g \times L_{g3}$, this will attenuate near-field coupling.

In order to fabricate the antenna, the dimensions are optimized for both isolation and compact size. The fabricated proposed antenna is shown in Figure 2. Table 1 shows the dimension of the proposed antenna.

**Mutual Coupling Reduction Performance**

Three possible phenomena can attribute to mutual coupling in patch antenna elements including near-field coupling, far-field coupling, and surface-wave coupling [4].

In the first step, surface wave which is covered enough in the literature, is analyzed and eliminated. Surface waves are excited in the patch antenna and they travel between the substrate and the ground plane [15]. They cannot exist where there is no dielectric slab to support them. Moreover, higher order modes of surface waves are excited as the substrate thickness or permittivity increases. Therefore, an approach to increase isolation due to surface wave is reducing the dielectric thickness or permittivity. Here, a thin thickness of $0.787\ mm$ is selected and Rogers substrate with relative permittivity of 2.2 is used.

Through the following, one can analyze the level of surface wave propagation. If the dielectric in between the antenna elements is removed, the air in between the patch elements eliminates the surface wave propagation. Accordingly, any difference between the $S_{12}$ values of the antenna structure with and without the dielectric implies surface wave effects, for an illustration of this surprising behavior, see Figure 3. Nevertheless, as shown in Figure 3, mutual coupling still exists in between of the antenna elements. Accordingly, this coupling may contribute to near-field and far-field coupling.

In the second step, near-field and far-field coupling are analyzed and eliminated. Far field radiation pattern of the antenna of Figure 4(a) is omnidirectional in the horizontal plane, if the dielectric is replaced by air [15]. Since, dielectric does not exist there is no dielectric atoms to become polarized and hence affecting the radiation pattern. This is shown in Figure 5. Replacing air by a dielectric constant of 2.2 affects the radiation in the horizontal direction. Therefore, antenna structure with a preferable low dielectric constant should be engineered in a way that there exists little radiation in horizontal direction to side antenna element. Figure 5 shows the x-y plane radiation pattern of a two element array. The solid line is related to the pattern of the structure with air as substrate. This shows an omnidirectional pattern. If air is replaced by a dielectric substrate, Figure 5 dashed line, shows that the related radiation pattern has a peak towards the neighboring patch element. Through modifying the antenna structure, one can manipulate the radiation pattern in such a way that it would have null towards the neighboring patch element. Thus, the coupling between the two patches would be reduced.

In this paper, through bending the feed lines, placing a slot in the ground beneath the feed lines and adding a stub in between the patch elements one can obtain the improved isolation. Through simulation one can justify the above. Figure 6 shows the S-parameters demonstrating the concept introduced.

To better understand the antenna behavior one can view the current distribution, Figure 4. The current distribution results show that by introducing the mechanism proposed, once an element is fed, the neighboring element is hardly excited, confirming that mutual coupling is reduced between the elements. The related near field distribution of the proposed antenna structure is shown in Figure 7. From Figure 7 is crystal clear that how near-field coupling changes by adding various parts to the proposed UWB MIMO antenna. It is worth to mention that the modifications done to the array structure can have a slight effect on impedance matching of the antenna.

The S-parameters for different structure of the proposed UWB MIMO antenna is plotted in Figure 6. It is obvious from Figure 6 that using h-shaped slot or long ground stub between the two antenna elements independently can decrease $S_{12}$ by $10\ dB$ due to mutual coupling. This demonstrate the repeatability of the proposed method to apply to any different structure.

**Results and Discussion**

*S-parameters*

The proposed MIMO antenna has been fabricated and tested. Computer simulation and measurement have been used to study the antenna performance. The simulated and measured S-parameters are in a good agreement as shown in Figure 8. The results in Figure 8 show an operation frequency from 3 to 12 $GHz$ and also with the introduced technique a mutual coupling of less than $-25\ dB$ is achieved. Also a return loss of better than $10\ dB$ all over the bandwidth are due in large to the fact that electric field is quadratically related to current distribution near the edges which is not considered in the HFSS simulation.

*Radiation pattern*

The simulated and measured radiation patterns of the proposed MIMO antenna in the x-y (H-plane) and y-z (E-plane) planes at 3.5, 5 and 8.5 $GHz$ are given in Figure 9. Patterns are obtained with one element fed and the other impedance matched.

At the frequency of 3.5 and 5 $GHz$, from Figure 9(a) and (b), it is observed that H-plane pattern is omnidirectional and E-plane is also very stable with changes in frequency. At the high frequency of 8.5 $GHz$ (Figure 9(c)), the H-plane pattern is sort of omnidirectional but E-plane pattern is not very stable with changes in frequency.

*Comparison of reported antenna structures*

The analysis of diversity performance has been also carried out on the proposed antenna which are shown in Figure 10 and 11. Envelope correlation of lower than 0.001 and capacity loss of less than 0.4 $bps/Hz$ was achieved. A comparison between the results reported in the available articles in the literature and the results obtained in the present work is given in Table 2. The proposed antenna of the present work has $S_{12} \leq -25\ dB$ throughout the bandwidth from 3.1 to 10.6 $GHz$ with the parallel port position and simulated radiation efficiency of higher than 95%. These characteristics make the proposed antenna a good candidate for small size UWB applications.

**Conclusion**

A novel technique to improve isolation between two closely-spaced antenna elements was introduced. The proposed antenna with a small size of 30.1×20.5 $mm^2$ and a very high isolation with $S_{12}$ less than -25 $dB$ was presented. The technique improved isolation significantly especially at low frequencies which is hard to achieve. This reduction was also demonstrated by using simulation of current distribution. The proposed technique can be utilized easily in other multiple antenna structures. The simulated radiation efficiency of higher than 95% with a stable omnidirectional pattern at all three shown frequencies was also obtained. These results make the proposed antenna a good candidate for portable UWB application.

Table 1. Dimensions of the proposed antenna in (mm)

| $W_r$ | $L_r$ | $W_m$ | $L_m$ | $L_f$ | $L_{s1}$ | $L_{s2}$ | $L_g$ | $D_f$ |
|---|---|---|---|---|---|---|---|---|
| *10* | *8* | *2.1* | *0.7* | *9.35* | *4* | *4.6* | *7.4* | *4.1* |
| $L$ | $W$ | $W_s$ | $W_{s1}$ | $W_a$ | $L_{f1}$ | $L_s$ | $L_{g3}$ | $L_{g1}$ |
| *20.5* | *30.1* | *3* | *0.7* | *19.2* | *3.5* | *19* | *0.5* | *1* |

Table 2. Comparison of the present work and those of previously reported works

| Reference | Port position | Size comparison (Proposed/ Reference) | Bandwidth ratio of Proposed/ Reference $S_{12} < 20\ dB$ |
|---|---|---|---|
| [15] | vertical | 42.44% | 43% |
| [16] | parallel | 69.90% | 28% |
| [17] | parallel | 109.98% | 68% |
| [18] | parallel | 44.94% | 52% |
| [19] | vertical | 62.91% | 27% |
| [20] | parallel | 37.5% | 45% |
| Present work | parallel | 100% | 100% |

Figure 1.  Geometry of proposed antenna (▬ top layer and ▬ bottom layer).

Figure 2.  Photograph of top and bottom view of the fabricated antenna.

Figure 3. Effect of thickness and air separation on surface wave causing mutual coupling. (Numbers correspond to thickness of the substrate in millimeter).

Figure 4. Current distribution of the proposed UWB MIMO antenna at 4.5 $GHz$. (a) The basic structure with no mutual coupling reduction technique. (b) Bending the feedline which results to reverse induced current. (c) h-shaped slots are applied to reduce near-field coupling. (d) The grounds are separated from each other. (e) A long strip is applied without h-shaped slots. (f) A long strip is applied with h-shaped slots.

Figure 5. Horizontal radiation pattern of the patch antenna for red:▬ configu-ration of Figure 4(a) and suspended in air; blue:··· configuration of Figure 4(a) with substrate; green:▬ configuration of Figure 4(f) with substrate. (right port excited at 4.5 $GHz$, the other matched.)

Figure 6. S-parameters of the different structures of the proposed UWB MIMO antenna (a) S12 (dB) (b) S11 (dB).

Figure 7. Near-field distribution of the proposed UWB MIMO antenna at 4.5 $GHz$ (a) The basic structure with no mutual coupling reduction technique (b) Bending the feedline to separate antenna elements more efficiently (c) H-shaped slots and ground separation are etched to improve isolation (d) A long strip is applied to prevent near-field coupling to the side antenna element.

Figure 8.  S-parameters of the proposed MIMO antenna.

Figure 9. Measured E-plane and H-plane radiation patterns of the proposed two-element MIMO antenna at (a) 3.5, (b) 5 and (c) 8.5 $GHz$ (red:▬ co simulated; blue:··· co measured; green:▬ cross simulated; pink:▬· cross measured)

Figure 10. Envelope correlation of the MIMO using S-parameter [21] (Diversity configuration of Figure 1)

Figure 11.  Measured and simulated capacity loss. (Configuration of Figure 1)